# Threadbare Parallel Port DAQ Card

Sumit Ghambir and P.Arun
Department of Physics and Electronics
S.G.T.B. Khalsa College
University of Delhi, Delhi 110 007
INDIA

*Abstract*

*A very simple Data acquisition card was developed in our under-graduate class for interfacing with the parallel port of a standard computer. It was developed with the intention to enable students to do various physics experiments with the computer sparing them from knowing the details of the electronics required. The article is a summary of our experience and ends with a request for suggestions to improve the circuit without losing its simplicity.*

The personal computer (PC) has two windows to communicate with the outside world, namely the serial and the parallel port. While the PC talks in binary (1's and 0's) the outside world's language is analog. Hence a translator is required for the PC to communicate with the outside world. This translation is done by an interfacing card, which is also popularly known as a "Data Acquisition Card" (DAQ). The heart of a DAQ is the analog to digital converter (ADC) chip. Depending on whether ADC's output is transferred one at a time or all at once, one uses the parallel or serial port respectively. Interfacing cards and DAQ systems are now commercially available and the in the western world introductory level interfacing cards are easily affordable to hobbyist. Various articles on how to design an interfacing card are available in popular electronics magazines and numerous web sites [1-5]. Infact a search on the Internet tends to swamp a person with too much information.

In this paper we record our experience of designing and performance analyzing of an interface card based on ADC0809, an eight bit ADC chip. We have made an attempt to develop a threadbare design so that the circuit would (i) be affordable, (ii) easy to replicate and also (iii) weeding basic idea from the clutter of information so as to minimize the technical information required to get started. Once the fundamentals are acquired a better design with more complicity can be developed. On the onset we would like to justify our selection of the ADC0809 chip and connecting it to the parallel port. The ADC0809 chip [6] is an eight bit analog to digital converter chip which gives the output as a parallel steam of eight bits along with signals called "Start Of Conversion" (SOC) and "End of Conversion" (EOC) which helps in hand-shaking (synchronizing) events with the PC. It is also an eight channel ADC i.e. it can convert eight different analog signals depending on which channel is selected. This feature is only used if the experimenter is converting his PC to a dual/multi-channel oscilloscope. Similarly, we selected the PC's parallel port since all eight bit output of the ADC can be read simultaneously making the programming simpler as compared to the serial port. Thus, the parallel port presents itself as an inexpensive platform for low frequency data acquisition.

The parallel port was developed as a communication port to the printer. Thus, it is referred by the more common and popular name of the "printer port". The computer sends data to the printer hence early printer ports had very few input pins (only 5). Technically, these ports were called the "Standard Parallel Port" (SPP). Intel has phased out the SPP, however even today computers based on AMD micro-processor and their chip-set have SPP. Interfacing cards however can be made for SPP also [7], however, they require additional hardware and signals for synchronization. Since 1994, parallel ports are bi-directional (PS) allowing eight pins either to act as input or output ports and more importantly these have been made TTL compatible. Parallel ports have further evolved and are presently classified as "Enhanced Parallel Port" (EPP) and "Enhanced Compatible Port" (ECP), details of which can be found in Jan Axelson's book [8]. This project was implemented using the parallel port in the PS mode.

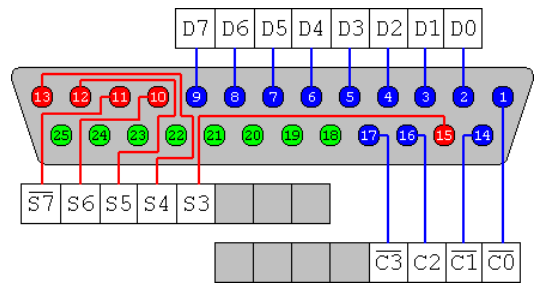

| Pin No | Signal name | Direction | Register - bit | Inverted |
|---|---|---|---|---|
| 1 | nStrobe | Out | Ctrl-0 | Yes |
| 2 | Data0 | In/Out | Data-0 | No |
| 3 | Data1 | In/Out | Data-1 | No |
| 4 | Data2 | In/Out | Data-2 | No |
| 5 | Data3 | In/Out | Data-3 | No |
| 6 | Data4 | In/Out | Data-4 | No |
| 7 | Data5 | In/Out | Data-5 | No |
| 8 | Data6 | In/Out | Data-6 | No |
| 9 | Data7 | In/Out | Data-7 | No |
| 10 | nAck | In | Status-6 | No |
| 11 | Busy | In | Status-7 | Yes |
| 12 | Paper-Out | In | Status-5 | No |
| 13 | Select | In | Status-4 | No |
| 14 | Linefeed | Out | Ctrl-1 | Yes |
| 15 | nError | In | Status-3 | No |
| 16 | nInitialize | Out | Ctrl-2 | No |
| 17 | nSelect-Printer | Out | Ctrl-3 | Yes |
| 18-25 | Ground | - | - | - |

*Figure 1:* *The pin layout of a parallel port and their significance.*

Anyone trying to replicate this project is advised to enter their computer's BIOS and verify their LPT (that's how your computer refers to the parallel port) is in PS/ bidirectional mode. While at it, also note the address of the LPT port. This address would be referred as the BASE address throughout the article.

The parallel port, visible at the back of your computer, has 25 pins with each pin having a role of its own. Figure 1 shows the pin position and list their grouping and role. Before proceeding to the circuit, we list here some important information of the parallel port. The 25 pins of the parallel port are grouped into Data, Control, Status and ground lines. Colloquially these lines are also called Data, Control and Status ports. The lines are connected to their corresponding registers inside the computer. So by manipulating these registers by programming, one can easily read or write to parallel port using languages like "C" and "BASIC" etc.

**1. Data port** (8 pins from 2-9):

For example the Data register is connected to Data lines, Control register is connected to control lines and Status register is connected to Status lines. So what ever you write to these registers, will appear in corresponding lines as voltages. Similarly, whatever you give as input at the parallel port (+5volts or 0volts) can be read from these registers (as 1's and 0's respectively).

The Data Port or Data Register is simply used for outputting/inputting data on the Parallel Ports' data lines (Pins 2-9). This register is default a write only port. If you read from the port, you should get the last byte sent. However if your port is bi-directional, you can receive data on this address. The Base Address shown in the BIOS is the address of the DATA port and in our PC it is 888D (D implies address is in decimal).

The STATUS port's address is BASE+1 (i.e. 889D) and that of the CONTROL port is BASE+2 (890D). Most of the computers list their BASE address in hexadecimal that of course can be converted to decimal.

**2. Status port** (5 pins from 10 to 13 &15)

The Status Port is a read only port. Any data written to this port by the programmer for outputting would be ignored. The Status Port is made up of 5 input lines (Pins 10,11,12,13 & 15). Bit 7 of the status register is active low i.e. if bit 7 happens to contain logic 0, there would be +5v at pin 11.

| Read/Write | Bit No. | Properties |
|---|---|---|
| Read Only | Bit 7 | Busy |
| | Bit 6 | Ack |
| | Bit 5 | Paper Out |
| | Bit 4 | Select In |
| | Bit 3 | Error |
| | Bit 2 | IRQ (Not) |
| | Bit 1 | Reserved |
| | Bit 0 | Reserved |

*Figure 2: Details of the Status port.*

**3. Control port** (5 pins 1, 14, 16 & 17)

The Control Port was intended as a write port only. When a printer is attached to the Parallel Port, four "controls" are used. These are Strobe, Auto Linefeed, Initialize and Select Printer, except for Initialize all of which are inverted. Since the printer isn't supposed to send any signal to the computer, default the Control port is an output port. However, the Control port can be programmed to work as an inputs port.

| Read/Write | Bit No. | Properties |
|---|---|---|
| Read/Write | Bit 7 | Unused |
| | Bit 6 | Unused |
| | Bit 5 | Enable Bi-Directional Port |
| | Bit 4 | Enable IRQ Via Ack Line |
| | Bit 3 | Select Printer |
| | Bit 2 | Initialize Printer (Reset) |
| | Bit 1 | Auto Linefeed |
| | Bit 0 | Strobe |

*Figure 3:* *Details of the Control port.*

Now we are in a position to discuss how to connect the various components required in our circuit and explain it more knowledgeably. We have already stated that the ADC requires a SOC pulse from the computer to start the process of conversion (this of course also synchronizes the two). Since the control port can be used as an output port Bit 3 (of the control port) i.e. pin 17 (of the parallel port) is used for giving Start of Conversion pulse to ADC. Similarly, since the Status Port is an input port; we use it to receive the End of Conversion pulse that would then tell the computer that the ADC has completed the conversion process and the data is available for the Data port to read and save in the computer. Figure 4 pictorially explains the above-mentioned idea. We now briefly discuss how programming generates the SOC.

*Generating the SOC Pulse*

To generate SOC pulse the output of pin C3 has to go high (+5v) for a time period greater than 20ns and then return to ground level. Since the output of C3 is inverted to whatever is written in the third bit of the control register, we give the following instructions (these instructions are of turbo basic, a variant of BASIC)

Out 890, 0

Delay 0.05

Out 890, 8

The Delay command introduces a high pulse of 0.05sec (50ms).

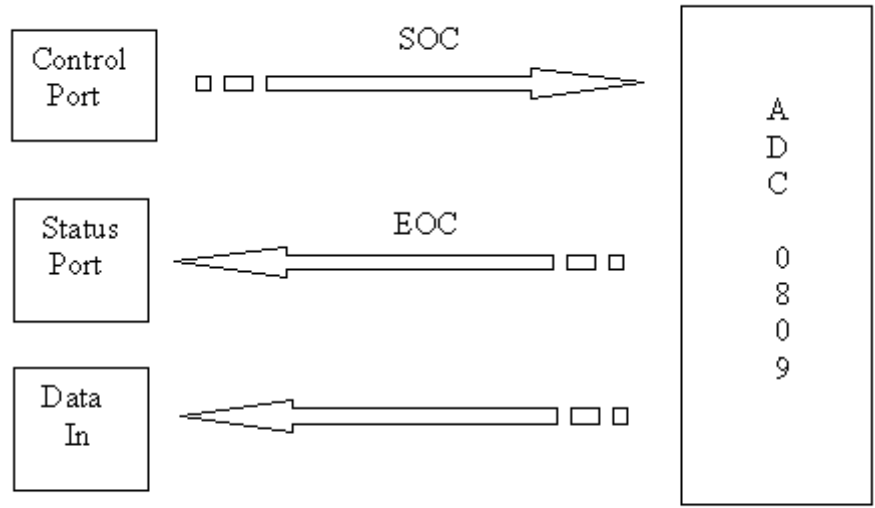

*Figure 4:* *Details of using the various ports to handshake with the ADC0809*

### Reading the EOC Pulse

The ADC gives EOC pulse (signal goes low on completion of conversion) to the parallel port (Status port, pin S3). The Status port works as an input port only when C0 is high. Since output of C0 is also inverted, which writing to the control register the Least Significant Bit (LSB) should be made low (bit should be zero). This was invariably the case when for SOC we sent 0 and 8 to the control register. To receive the EOC pulse we read the status port, using the INP command of TB. The syntax is

a% = Inp (889)

where 889 is status port address and on its execution all the eight bits of the status register is saved in a set variable a%. The third bit is then checked for 1 or 0. A zero implies the arrival of EOC.

The ADC0809 is quite a developed chip with (Address Latch Enable) ALE pin for latching the address of the channel selected. Since we plan to use only a single channel, all the address pins are grounded (Channel 0 is selected) and ALE is tied to SOC. Also, to tri-state the data bus in a complex peripheral circuit, an Output Enable (OE) pin is given. Since we are only using one chip in our circuit, we allow it to demand the bus at all time by keeping OE high by connecting it to $V_{cc}$, and $(+)V_{ref}$ pin which are given +5v. $(-)V_{ref}$ is grounded.

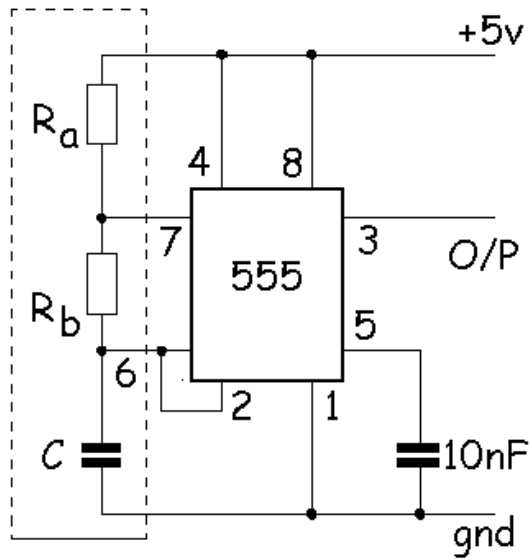

*Figure 5: Circuit diagram of IC555 in Astable mode. The circuit was used as a clock for our interfacing card.*

The clock pulse required for the ADC0809 was given by an IC555 in astable mode [9] (see figure 5). The clock frequency that can be supplied to ADC0809 is between 200-640KHz. We designed our clock for frequency of 227.27 KHz. The output lines of ADC0809 (i.e. pins D0-D7) were given to the buffer 74LS244 (see figure 6) whose output was connected to the PC's parallel port by a 25-way-male connector. The buffer 74LS244 is placed between the output of ADC and parallel port to protect the computer from any excess current flow due to poor circuit designing or in the eventuality the ADC is destroyed. The buffer chip consists of eight operational amplifiers to who's input the data bits from ADC are connected and their outputs are connected to the parallel port.

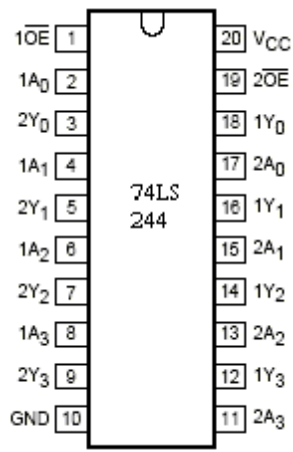

*Figure 6:* *Pin layout of IC74LS244, an octal buffer chip.*

Technically the octal buffer 74LS244 is a typical example of tri-state buffer, it also known as a line driver or line receiver. It has two groups of four buffers with non-inverted tri-state output and the buffers are controlled by two active low Enable lines {1OE & 2OE both inverted}. Until these lines are enabled, the output of the drivers remains in high impedance state disconnecting the circuits at the input and output. A PCB (Figure 7) was designed by us using SPRINT software and then fabricated for less than Rs100/-.

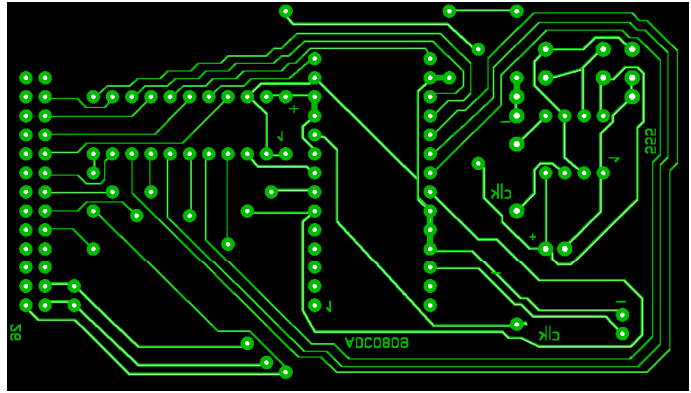
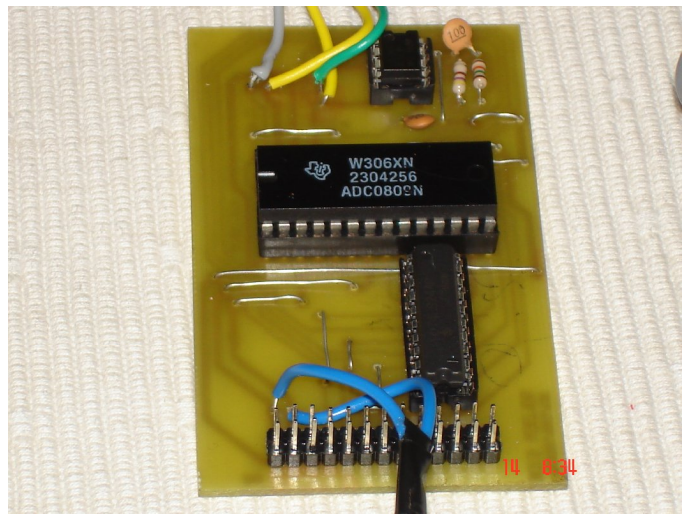

***Figure 7:*** *Designed PCB on computer and as seen after assembling the circuit. Actual size of the PCB is 9cm x5cm.*

To test the interfacing card, different waveforms were given at the input of ADC0809 from a function generator. Before doing so, we have to be very careful and signal condition the input. That is, made sure that the input waveform's peak-to-peak voltage was between 0 to +5volts and that its frequency was low in Hz.

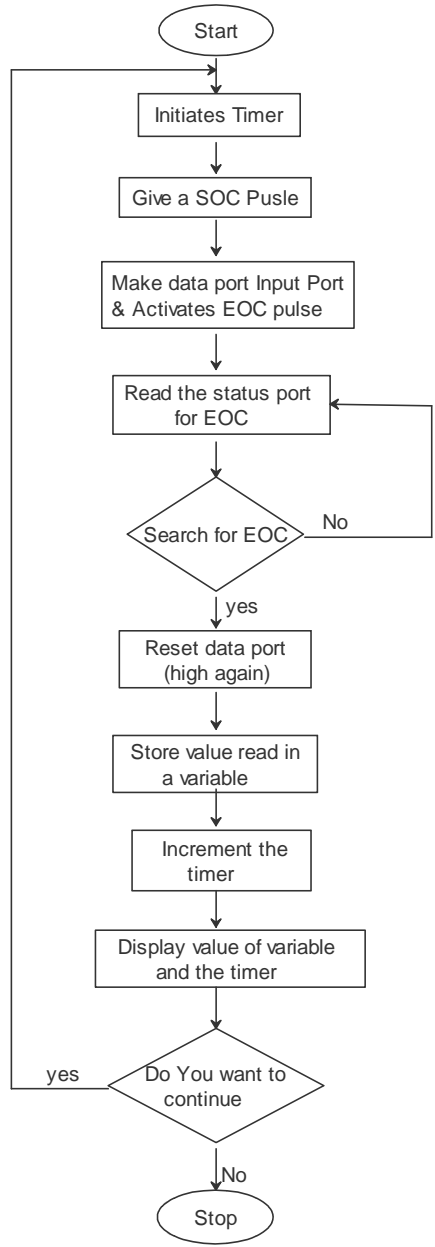

*Figure 8: Flowchart for program to continuously capture data from designed interfacing card and store data as time and instantaneous voltage.*

To implement this flow chart as a program of instructions for the computer, we decided to use Turbo-basic as our programming language. Turbo-basic (TB) is a High-Level Language with English like commands. This feature makes TB very elementary and easy to learn. Moreover, TB comes with a library function called MTIMER (micro-timer), which theoretically allows to time events with a 1-microsecond resolution. Using this we can work out the frequency of the waveform input to our interfacing card.

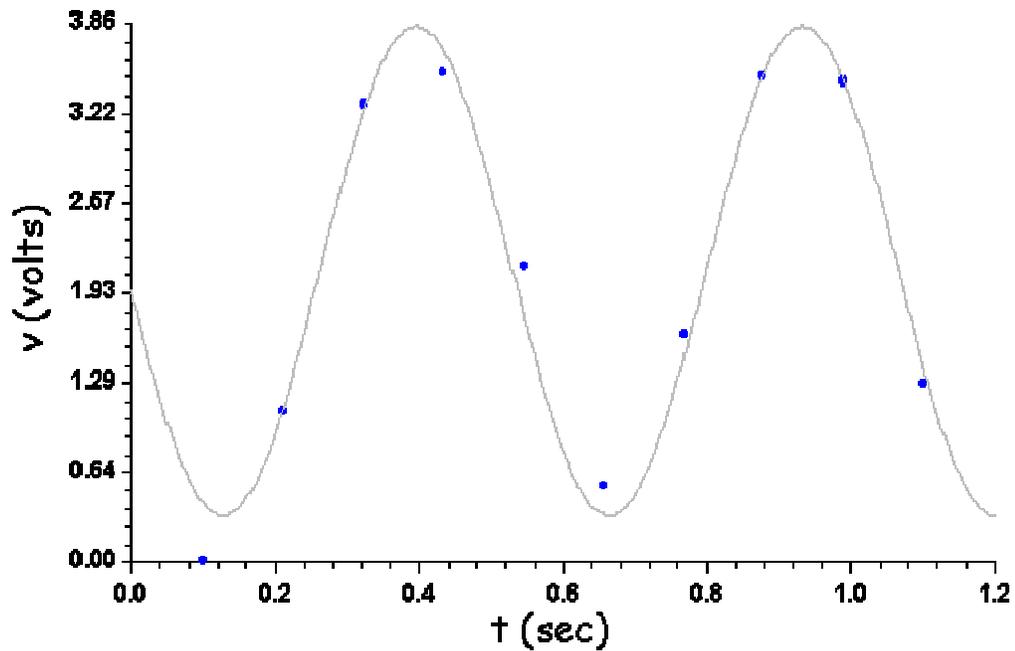

*Figure 9:* *Display of captured data of an input sine wave.*

We can also plot the saved data in a voltage versus time graph to get a display similar to that seen on an oscilloscope (see figure 9, detail of which would be discussed later). However, we found that Mtimer was not a reliable timer in the microseconds regime but was working very reliable in the milliseconds regime. This crippled our ability to work with very high frequencies but technically we still could work and capture data signals with frequencies in KHz. It should be noted here that we couldn't access the PC's parallel port by user written programs with Microsoft operating systems beyond WINDOWS NT/2000/XP. For systems with these operating systems you either have to develop a driver for your interfacing card or use Visual Basic with supporting "INPUT32.DLL" file for WINDOWS NT/2000/XP.

Figure 9 is a plot of the data stored as an ASCII file and plotted using the software "Wgnuplot". While the trend of the sine wave is evident, it was impossible to quantify the wave confidently. That is, one couldn't comment on $V_m$ or the frequency confidently. For this software "Curxprt" was used for curve fitting. The software returns the best fitting curve, which is shown, as the continuous line in fig 9. We had written and tested many versions of data logging programs with some modifications in each version. Figure 10 gives our final program that received data at the fasted rate and with good reproducibility. As can be seen from fig 9, five data points are collected in a single cycle. Thus, the sampling rate is 10 samples per second or 1 reading in 0.1sec. This immediately rules out data acquisition in KHz as limited by the Mtimer function of TB. Infact the sampling rate shows our circuit can only work reliably in frequency ranges of Hz.

After many trials we found that our circuit gives erratic results if data is collected at rates faster than 1 reading in 0.05sec. Thus a delay of 0.05sec has to be incorporated between two readings. This is done by inserting a delay of 0.0275sec before looping (see program). The remaining delay of 0.0225sec is presented by the time taken for program implementation, i.e. time taken by computer to execute the various commands given. If we call the total delay (0.05sec) the Reset delay, we may write

**Reset Delay = Program Delay + Insert Delay**

where Insert Delay is the delay inserted in program using delay command (0.0275sec in our case). The program delay is due to the complicity of a high level language and can be minimized by resorting to programming in Assembly language. We tested this circuit on 8085 microprocessor training kit [10]. While the program delay was bought down we saw that the insert delay had to be increased such that the reset delay remained constant. This implies that if a second SOC is given to the ADC0809 0.05sec before the lapse of the first EOC, the behavior of the chip becomes unreliable. This in turn tells us that with a reset time of 0.05sec our sampling rate becomes (1/0.05 Hz =) 20 Hz. If we demand 10 samples for each cycle of input waveform, the maximum frequency we can reliably input and recreate is (20/10 Hz =) 2Hz.

The datasheet of ADC0809 says SOC can be given immediately after EOC and suggests connecting EOC to SOC for maximum conversion rate, which technically would allow data acquisition of 10KHz signals. Previous published works are silent on this as their programs did not have looping for continuous streaming of data. We were not able to resolve this problem of reset time and hence invite feedback from anyone with plausible solutions. However, our design was successful in teaching us the fundamentals behind interfacing and allowed us to do slow event physics experiments.

## Acknowledgements

We would like to express our gratitude to the lab staff of S.G.T.B. Khalsa College and also U.G.C. for the assistance (No.F.6-1(25)/2007(MRP/Sc/NRCB).

```
$include "c:\tb\intrpt"
dim time(300), voltage(300), ii(300)
cls
input "Give test frequency in hz",hz
hz=1/hz
?hz
hz=3*hz
?hz, hz/0.05
open "O",1,"c:\tb\f10hz.dat"
sum&=0
i=0
do
mtimer
'-------------------------SOC---------------------
out 890,8
out 890,0
delay .05
out 890,8
'-------------------------SOC---------------------"
do
b%= inp(889)
mb%= b% and 8
    if (mb%/8)=0 then
    time&=mtimer
    out 890,40
    c%= inp(888)
    sum&=sum&+time&
    i=i+1
  time(i)=sum&*1e-6
  voltage(i)=5*c%/255
  ii(i)=i
    end if
loop while mb%><0
delay 0.0275
loop while sum&*1e-6<hz
?"end"
for j=1 to i
print #1,time(j),voltage(j),ii(j)
next j
```

*Figure 10:* The Turbo basic program used along with our hardware for data acquisition.